\documentclass[prb,showpacs,twocolumn]{revtex4}

\newcommand\Ref[1]{(\ref{#1})}
\usepackage{amssymb}
\renewcommand\vec[1]{\ensuremath{\boldsymbol{#1}}}
\usepackage{amsmath}
\usepackage[dvipdfm]{graphicx}
\usepackage[dvipdfm]{color}
\usepackage{tikz}
\newcommand{\DeltaTheta}{\ensuremath{\delta\negthinspace\theta}}
\newcommand{\Deltatheta}{\ensuremath{\delta\negthinspace\varphi}}
\newcommand{\Deltaphi}{\ensuremath{\delta\negthinspace\phi}}
\graphicspath{{./newplots/}}

\begin{document}
\bibliographystyle{plain}
\title{Barnett Effect in Thin Magnetic Films and Nanostructures}
\author{Stefan Bretzel}
\author{Gerrit E. W. Bauer}
\affiliation{Kavli Institute of NanoScience, Delft University of Technology,
Lorentzweg 1, 2628 CJ Delft, The Netherlands}
\author{Yaroslav Tserkovnyak}
\affiliation{Department of Physics and Astronomy, University of California, Los Angeles, California 90095, USA}
\author{Arne Brataas}
\affiliation{Department of Physics, Norwegian University for Science and Technology, 7491 Trondheim, Norway}
\begin{abstract}
The Barnett effect refers to the magnetization induced by rotation of a demagnetized ferromagnet. We describe the location and stability of stationary states in rotating nanostructures using the Landau-Lifshitz-Gilbert equation. The conditions for an experimental observation of the Barnett effect in different materials and sample geometries are discussed. 
\end{abstract}

\maketitle
At the dawn of quantum mechanics, the Barnett\cite{Bar15,Bar35} effect|magnetization induced by rotation|confirmed that magnetization is associated with angular momentum. Furthermore, Barnett measured the gyromagnetic ratio of electrons in ferromagnets and the anomalous $g$ factor of the electron for the first time. The Barnett effect can be understood in terms of a rotating gyroscopic wheel, that aligns itself with the axis of rotation until a stationary state in the rotating frame of reference is achieved. Since angular momentum $\vec{L}$ is associated with magnetization $\vec{M}=-\gamma \vec{L}$, with $\gamma=g\mu_B/\hbar=g\vert e\vert/2m$ being the gyromagnetic ratio, mechanical rotation induces a net magnetization antiparallel to the axis of rotation. The torque acting on the magnetization in the rotating frame of reference is equivalent to a torque due to the presence of a gauge magnetic field
\begin{equation}
\vec{H}_{\mathrm{rot}}=-\gamma^{-1}\vec{\omega}\,.\label{gaugefield}
\end{equation}

There has recently been a renewed interest in the coupling of
magnetization with mechanical motion, for example in mechanically detected ferromagnetic resonance spectroscopy measurements.\cite{Ru92} A nano-magnetomechanical system consisting of a cantilever and a thin magnetic film shows
coupled magnetovibrational modes.\cite{Ko03, Wi06}  Furthermore, the nanomechanical current-driven spin-flip torque at the normal-metal/ferromagnet interface of a suspended nanowire has been detected.\cite{Zol08}

In Barnett's original experiments, rotation frequencies of $\omega\lesssim 500 \mathrm{\mbox{ }Hz}$ generated a change of the magnetic field of the order of $10^{-4}$~Gauss in macroscopic samples. Although in nanostructures detecting such small fields may become more challenging, a range of powerful techniques have recently been developed, which could be utilized for the purpose. 
To date, very small changes in the magnetization can be measured using the magneto-optical Kerr effect, Faraday spectroscopy, superconducting quantum interference devices (SQUID's) or Hall micromagnetometry.\cite{Ge97} Therefore, we present here a theoretical feasibility study of the Barnett effect in magnetic thin films and nanostructures. Our focus is the dynamics in magnetic
thin films and nanoclusters, which we study by means of the 
Landau-Lifshitz-Gilbert (LLG) equation for the magnetization vector $\vec{m}$:
\begin{equation}
\dot{\vec{m}}=
-\gamma\vec{m}\times\vec{H}_{\mathrm{eff}}+\alpha\left.\vec{m}\times\dot{\vec{m}}\right\vert_{\mathrm{Lat}}\,,\label{llg}
\end{equation}
where $\vec{H}_{\mathrm{eff}}$ is the effective magnetic field, $\vec{m}$ is the unit vector of magnetization and $\alpha$ the dimensionless damping constant. We can separate the dynamics caused by the rotation of the system as a whole from the dynamics in the rotating frame of reference by the transformation $\vec{m}=R(\phi)\vec{m}_R$ and $\vec{H}_{\mathrm{eff}}=R(\phi)\vec{H}_{\mathrm{eff}}^R$, where $R(\phi)$ is a unitary matrix describing the rotation by a time-dependent angle $\phi(t)$ around the axis of rotation  and $\vec{m}_R$ ($\vec{H}_{\mathrm{eff}}^R$) denote the magnetization (effective magnetic field) in the rotating frame of reference. The damping is caused by the magnetization motion relative to the lattice:
\begin{equation}
\left.\vec{m}\times\dot{\vec{m}}\right\vert_{\mathrm{Lat}}=R(\phi(t))\left(\vec{m}_R\times\dot{\vec{m}}_R\right)\,.
\end{equation}
In the rotating frame of reference Eq.~\Ref{llg} becomes 
\begin{equation}
\dot{\vec{m}}_R=\vec{m}_R\times\left(-\gamma\vec{H}_{\mathrm{eff}}^R+\omega\vec{e}_z+\alpha\dot{
\vec{m}}_R\right)\,.\label{llg_rot}
\end{equation}
In this derivation, we have tacitly assumed that the Hamiltonian transforms trivially under rotation, i.e. rotation only generates the gauge Zeeman field Eq.~\Ref{gaugefield} in the rotating frame of reference. 
[Note that if rotation stems from a rotating external field\cite{dAqu04-2,dAqu04} rather than the lattice, we would have to use a different form of damping, {\em viz.} $\vec{m}=R(\phi)\vec{m}_R$ in $\vec{m}\times\dot{\vec{m}}$. Then the right
hand side of Eq.~\Ref{llg_rot} contains an additional term
$\alpha\omega\vec{m}_R\times\vec{e}_z$ and the stationary states of
Eq.~\Ref{llg_rot} depend on the damping constant $\alpha$.]

Following Barnett,\cite{Bar15} we are
looking for stationary state solutions in the rotating frame of reference, i.e., solutions $\vec{m}_R$ for which $\dot{\vec{m}}_R=0$.  From Eq.~\Ref{llg_rot} it follows that the stationary states obey:
\begin{equation}
0=\vec{m}_R\times\left(-\gamma\vec{H}_{\mathrm{eff}}^R+\omega\vec{e}_z\right)\,.
\end{equation}
Here the magnetization in the lab frame of reference 
precesses around the axis of rotation ($z$-axis) at a fixed angle. 
We analyze the stability of the stationary states in
spherical coordinates, i.e., 
\begin{equation}
 \vec{m}_R=(\sin\theta\cos\phi, \sin\theta\sin\phi, \cos\theta)\label{mrparametrization}
\end{equation}
by linearizing the set of equations resulting from Eq.~\Ref{llg_rot} for small deviations $(\DeltaTheta, \Deltaphi)$ from the equilibrium (rotating-frame) positions $(\theta_n, \phi_n)$. When $\vec{H}_{\mathrm{eff}}=0$, e.g., in a spherical particle without crystal anisotropy, the stationary states are given by $\pm \vec{e}_z$. Clearly the
stationary state at $\vec{e}_z$ is unstable and 
$-\vec{e}_z$ is stable.

\begin{figure}
\includegraphics{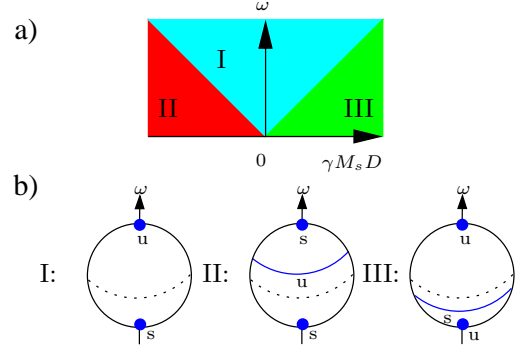}
\caption{(Color online) I, II and III indicate regions in the $(\omega,\gamma M_s D)$ plane with stable and unstable stationary states located at $\mp\vec{e}_z$, respectively, (region I), stable stationary states at $\pm\vec{e}_z$ and unstable stationary states located at a fixed angle $\theta=\arccos(-\omega/\gamma M_s D)$ in the upper half plane (region II) and stable stationary states located in the lower half plane and unstable stationary states at $\pm\vec{e}_z$ (region III).}

 \label{fig:cartoon}
\end{figure}

For a film with free energy $F=DM_z^2/2$, i.e., $D>0$ refers to a easy-plane magnetization and $D<0$ to an easy-axis magnetization parallel to the axis of rotation, $\vec{H}_{\mathrm{eff}}$ is given by $\vec{H}_{\mathrm{eff}}=-M_s\mathrm{diag}\{0,0,D\}\vec{m}$, where $\mathrm{diag}\{\dots\}$ refers to a diagonal matrix with entries $0$ and $D$ and $M_s$ refers to the saturation magnetization. Without limiting generality, we consider the case that $\omega>0$. 
By using Eq.~\Ref{mrparametrization} in Eq.~\Ref{llg_rot}:
\begin{equation}
 \dot{\theta}=-\alpha\dot{\phi}\sin\theta =\frac{\alpha}{1+\alpha^2}\left(\omega
+\gamma M_\mathrm{s}D\cos\theta\right)\sin\theta\,.\label{inplanethetaeq}
\end{equation}
Thus, the stationary states are given by $\sin\theta_1=0$ and, if $\omega\leq \gamma M_\mathrm{s} \vert D\vert$, by
\begin{equation}
 \cos\theta_2=-\frac{\omega}{\gamma M_\mathrm{s}D}\,.\label{additionalstates}
\end{equation}
These fixed points do not depend on the coordinate $\phi$ due to the axial symmetry. 
For small deviations $\DeltaTheta$ from $\theta=0,\pi$, the linearized Eq.~\Ref{inplanethetaeq} yields
\begin{equation}
\delta\negthinspace\dot{\theta}=\frac{\alpha}{1+\alpha^2}\left(\gamma M_sD\pm \omega\right)\DeltaTheta\,,\label{linearizedtheta0pi}
\end{equation}
where the $+$ ($-$) sign refers to the steady state at $\theta=0$ ($\theta=\pi$). For $\omega<\gamma M_\mathrm{s}\vert D\vert$ there are additional stationary states $\cos\theta=-\omega/\gamma M_\mathrm{s}D$. For small deviations $\DeltaTheta$ from $\arccos(-\omega/\gamma M_\mathrm{s}D)$ the linearized Eq.~\Ref{inplanethetaeq} reads
\begin{equation}
\delta\negthinspace\dot{\theta}=\frac{\alpha}{1+\alpha^2}\frac{\omega^2-\gamma^2 M_\mathrm{s}^2D^2}{\gamma M_\mathrm{s} D}\DeltaTheta\,.\label{linearizedthetainbetween}
\end{equation}
We can now identify three different regimes: $\omega>\gamma M_s\vert D\vert$, $\omega<\gamma M_s\vert D\vert$ and $D<0$ or $D>0$ (regions I, II and III in Fig.~\ref{fig:cartoon}a). If $\omega>\gamma M_s\vert D\vert$, i.e., region I in Fig.~\ref{fig:cartoon}a, one sees from Eq.~\Ref{linearizedtheta0pi} that the stationary state $\theta=0$ ($\theta=\pi$) is unstable (stable). See cartoon I in Fig. \ref{fig:cartoon}b. When $\omega<\gamma M_s\vert D\vert$, additional steady states given by Eq.~\Ref{additionalstates} exist. If also $D<0$ (region II in Fig.~\ref{fig:cartoon}a), i.e., easy axis anisotropy, it follows from Eq.~\Ref{linearizedtheta0pi} that $\theta=0,\pi$ are stable and from Eq.~\Ref{linearizedthetainbetween} that $\cos\theta_2=-\omega/\gamma M_sD$ are unstable stationary states (see cartoon II in Fig.~\ref{fig:cartoon}b). However, if $D>0$ (region III in Fig.a), i.e., easy plane anisotropy, according to Eqs.~\Ref{linearizedtheta0pi} resp. \Ref{linearizedthetainbetween} $\theta=0,\pi$ are unstable and $\cos\theta_2=-\omega/\gamma M_s D$ are stable stationary states (see cartoon III in Fig.~\ref{fig:cartoon}b).

\begin{figure}
\includegraphics{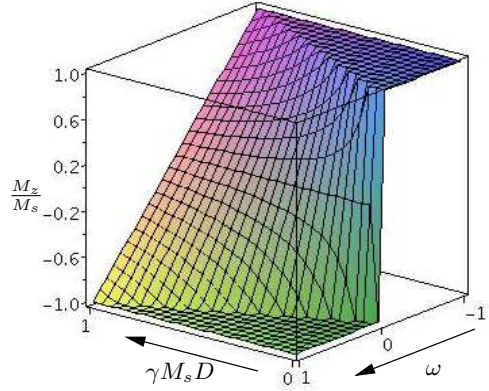}
\caption{(Color online) The $z$ component of the magnetization for the easy-plane configuration, i.e., $D>0$, in the rotation frequency $\omega$ vs. anisotropy field $\gamma M_\mathrm{s}D$ plane. Both $x$ and $y$ axes are scaled to have the same dimensions. }
\label{summary_stab_analysis}
\end{figure}

To summarize, in a system with in plane magnetization, i.e., $D>0$, the stable stationary states acquire a $z$ component by rotation. The rotation acts like a magnetic field along the magnetic hard axis. Fig. \ref{summary_stab_analysis} shows the $z$-component (component along the axis of rotation) of the magnetization in the stationary state in the $\omega$ vs. $\gamma M_s D$ plane. In this regime the magnetization displays a hysteresis loop when $\omega$ is cycled. The larger $\gamma M_s D$, the slower the transients become.

Limit cycles do not exist, since when $\omega$ is constant, we find for the time-derivative of the free energy $\dot{F}\gamma/M_s=-\alpha(\dot{\vec{m}}_R)^2$. In other words, the magnetization approaches its stationary state.

When the axis of rotation no longer coincides with the anisotropy axis of the crystal, the rotational symmetry around the axis of rotation is broken. As a consequence, only a finite number of fixed points exists. For an autonomous system on the unit sphere such as the LLG equation with time-independent effective field, it follows from the Poincare index theorem\cite{Per91,BSM01} then that the number of (un)stable fixed points minus the number of saddles must be equal to two.
A magnetic needle along the $y$-axis, i.e., $\vec{H}_{\mathrm{eff}}=M_s\mathrm{diag}\{0,D,0\}\vec{m}$ 
spun around the $z$ axis exhibits four stationary states when $\omega<\gamma M_\mathrm{s}\vert D\vert$: $\theta_{1,2}=0,\pi$ and $\cos\theta_{3,4}=-\omega/\gamma M_\mathrm{s}D$, $\cos\phi_{3,4}=0$. If $D>0$, then $\theta_1=0$ is an unstable and $(\theta_{3,4},\phi_{3,4})$ stable stationary states whereas $\theta_2=\pi$ is a saddle point. In the opposite case, i.e., $D<0$, $(\theta_{3,4},\phi_{3,4})$ are unstable and $\theta_2=\pi$ is stable, whereas $\theta_1=0$ is a saddle point.

For typical magnetic materials, the critical frequencies to fully rotate the magnetization from in-plane to perpendicular-to-plane orientation are inaccessibly high: $\omega\sim 
200\mathrm{\mbox{ }GHz}$ for permalloy with $M_s\sim 1000\mathrm{\mbox{ }emu}/\mathrm{cm}^3$ and $D\sim 4\pi$, 
 and  $\omega\sim 4\mathrm{\mbox{ }GHz}$ for a GaMnAs film\cite{Liu05} with $M_s\sim 15\mathrm{\mbox{ }emu}/\mathrm{cm}^3$ and $D\sim 4\pi$. However, to identify the Barnett effect, it is sufficient to observe
small changes in the $z$ component of the magnetization: $M_z=-\omega/\gamma D$.
For example, in metals polar magneto-optic Kerr spectroscopy is reported to be sensitive to magnetic moment changes down to $\sim  10^{-15}\mathrm{\mbox{ }emu}$ at a spot diameter of $0.5\mbox{ }\mu\mathrm{m}$.\cite{Cor08} For a $10\mathrm{\mbox{ }nm}$  thick permalloy film ($D\sim 4\pi$) this corresponds to a change in the magnetization of $M_z\sim 1\mathrm{\mbox{ }emu}/\mathrm{cm}^3$ which is achieved by a rotation frequency of $\omega\gtrsim 200\mathrm{\mbox{ }MHz}$. A Kerr angle of $0.3\mathrm{\mbox{ }deg}$ has been measured when the magnetization of GaMnAs is fully aligned perpendicular to the axis of rotation by an external magnetic field.\cite{Liu05} Together with a reported angular resolution in polar Kerr measurements\cite{All03} of $\sim 10^{-4}\mathrm{\mbox{ }deg}$ this yields a required rotation frequency of a few MHz. However, since the cubic anisotropy is important in GaMnAs,\cite{Liu05} the above number serves as a lower bound for the frequency estimate.
The Barnett effect can be observed at lower spinning rates by choosing a material with small anisotropies. The perpendicular anisotropy in thin magnetic films can be tuned by the layer thickness to cancel the shape anisotropy.\cite{Shen97,Daa92,Zha94,Phil01}

The Barnett effect can be also used to move domain walls. Consider a wire along the $y$ axis, which contains a transverse Bloch wall in the $xz$ plane. When the wire is rotated around the $z$ axis, the Bloch domain wall moves with a velocity\cite{Wal74} $v=\lambda_w\omega/\alpha$, where $\lambda_w$ is the width of the transverse Bloch domain wall. For $\lambda_w\sim 100\mathrm{\mbox{ }nm}$ and $\alpha\sim 10^{-2}$ this yields $v\sim(10\mathrm{\mbox{ }m/s})\cdot (\omega/\mathrm{MHz})
$. 

It might be easier to observe the Barnett effect by vibration rather than rotation, but the mechanical vibration amplitude $\Deltatheta$ then becomes an
additional control parameter. The magnetization response is enhanced
when the harmonic vibration and FMR frequencies coincide. At this magnetopolariton
mode,\cite{Ko03} a $z$ component of the magnetization is excited in a needle in the
$xy$ plane that oscillates around the $z$ axis. Assuming a vibration amplitude $\Deltatheta$ (rad),
$M_z$ oscillates with an amplitude $M_s\Deltatheta/2\alpha$. 

In the ideal case of zero anisotropy only the temperature-induced thermal activation of the magnetization has to be overcome in order to observe a Barnett effect, which sets the lower bound on frequency according to $VM_s\omega\gtrsim \gamma k_B T$. For a spherical particle with diameter $d$ and saturation magnetization $M_s$ this yields a minimum frequency of about 500~MHz at $T=1\mathrm{\mbox{ }K}$, $M_s=10 \mathrm{\mbox{ }emu}/\mathrm{cm}^{3}$ and $d=10\mathrm{\mbox{ }nm}$. For a $10\mathrm{\mbox{ }nm}$ thick film with $M_s=10\mathrm{\mbox{ }emu}/\mathrm{cm}^3$  and area $A=1\mathrm{\mbox{ }}\mu\mathrm{m}^2$ with compensating form and crystal anisotropies, the required rotation frequency is about 25~kHz. 

In conclusion, we discussed the Barnett effect in magnetic nanostructures, which gives a handle to manipulate magnetization by mechanical means. We find that the rotation frequencies necessary to fully switch magnetizations in conventional materials are very high and beyond present experimental possibilities. However, the Barnett effect can be observed via partial magnetization of very soft materials, rotation-induced domain-wall motion, and vibrations close to magnetic resonance frequencies.

This work is part of the research programme of the ``Stichting voor Fundamenteel Onderzoek der Materie (FOM)'', which is financially supported by the ``Nederlandse Organisatie voor Wetenschappelijk Onderzoek (NWO)''.

\end{document}